\documentclass[pra,aps,twocolumn,showpacs]{revtex4}

\usepackage{times,epsfig,amssymb,amsfonts,amsmath}%\usepackage{epsfig}

\newcommand{\ket}[1]{|#1\rangle}
\newcommand{\bra}[1]{\langle #1|}
\newcommand{\proj}[1]{\ket{#1}\bra{#1}}

\begin{document}

\title{Entanglement monogamy and entanglement evolution
in multipartite systems}

\author{Yan-Kui Bai$^{1,2}$}\author{Ming-Yong Ye$^{1,3}$}\author{Z. D. Wang$^1$}
%\email{zwang@hkucc.hku.hk}
\affiliation{ $^1$ Department of Physics
and Center of Theoretical and Computational Physics, University of
Hong Kong, Pokfulam Road, Hong Kong, China\\
$^2$ College of Physical Science and Information Engineering and
Hebei Advance Thin Films Laboratory, Hebei Normal University,
Shijiazhuang, Hebei 050016, China\\
$^3$ School of Physics and Optoelectronics Technology, Fujian Normal
University, Fuzhou 350007, China}

%%%%%%%%%%%%%%%%%%%%%%%%%%%%%%%%%%%%%%%%%%%%%%%%%%%%%%%%%%%%%%%%%%%%%%%%%%%%%%%%
\begin{abstract}
We analyze the entanglement distribution and the two-qubit residual
entanglement in multipartite systems. For a composite system
consisting of two cavities interacting with independent reservoirs,
it is revealed that the entanglement evolution is restricted by an
entanglement monogamy relation derived here. Moreover, it is found
that the initial cavity-cavity entanglement evolves completely to
the genuine four-partite cavities-reservoirs entanglement in the
time interval between the sudden death of cavity-cavity entanglement
and the birth of reservoir-reservoir entanglement. In addition, we
also address the relationship between the genuine block-block
entanglement form and qubit-block form in the interval.
\end{abstract}
%%%%%%%%%%%%%%%%%%%%%%%%%%%%%%%%%%%%%%%%%%%%%%%%%%%%%%%%%%%%%%%%%%%%%%%%%%%%%%%%


\maketitle

\section{introduction}
 As an important physical resource, entanglement has widely
been applied to  quantum communication\cite{eke91prl,ben93prl} and
quantum computation\cite{ben00nat,rau01prl}. It is fundamental to
characterize entanglement nature of quantum systems, especially at a
quantitative level. Until now, although the bipartite entanglement
is well understood  in many aspects, the multipartite entanglement
is far from clear \cite{rev07qic} and thus deserves profound
understandings.  In many-body quantum systems, one of the most
important properties is that entanglement is monogamous, which means
quantum entanglement can not be freely shared among many parties. As
quantified by the square of the concurrence \cite{woo98prl}, a
three-qubit monogamy inequality was given by Coffman \emph{et al.}
\cite{ckw00pra} as
$\mathcal{C}_{A|BC}^2\ge\mathcal{C}_{AB}^2+\mathcal{C}_{AC}^2$.
Recently, its $N$-qubit generalization was made by Osborne and
Verstraete \cite{osb06prl}. Moreover, using some other entanglement
measures, similar monogamy inequalities have also established
\cite{yus05pra,hir07prl,fan07pra,jhh08pra,chi08jmp,kim09pra}.
However, in these monogamous relations, only the single party
partition $A_{1}|A_{2}A_{3}\ldots A_{n}$ is considered. Whether it
can be generalized to other partitions, such as two parties cut
$A_{i}A_{j}|A_{k}A_{l}\ldots A_{n}$, is still an open question to be
answered.

On the other hand, the entanglement dynamical behavior under the
influence of environment is also an important property of quantum
systems. This is because that, in realistic situations, quantum
systems interact unavoidably with the environment, and may lose
their coherence. It was reported recently that an entangled state of
two qubits interacting respectively with two local reservoirs would
experience disentanglement in a finite time, even if the coherence
is lost asymptotically \cite{zyc01pra,tyu04prl,alm07sci,lau07prl}.
This phenomenon is referred to as entanglement sudden death (ESD),
and has received a lot of attentions both theoretically and
experimentally (see a review paper \cite{tyu09sci} and references
therein).

Recently, L\'{o}pez \emph{et al.} analyzed the entanglement transfer
between two entangled cavity photons and their corresponding
reservoirs, and showed that the entanglement sudden birth (ESB) of
reservoir-reservoir subsystem must happen whenever the ESD of
cavity-cavity subsystem occurs \cite{clo08prl}. However, in this
process, \emph{whether there exists an entanglement monogamy
relation restricting the dynamical evolution} is awaited for further
studies. Moreover, in the time interval where both the cavity-cavity
entanglement and the reservoir-reservoir entanglement are zero, a
subtle issue \emph{where the initial entanglement really goes} is
yet to be resolved, although the nonzero cavity-reservoir
entanglement in this time window was pointed out.

In this paper, based on a new monogamy relation, the entanglement
dynamics of two cavities interacting with individual reservoirs is
studied. It is found that the genuine multipartite entanglement is
involved in the dynamical process. Particularly, at a quantitative
level, we show the initial cavity-cavity entanglement evolves
completely to the genuine four-partite entanglement in the time
interval between the ESD and the ESB. In addition, we also address
the property of the genuine multipartite entanglement which exhibits
in the block-block form under the bipartite two-qubit partition.

\section{Two-qubit residual entanglement and monogamy relations}
Let us first recapitulate the monogamy inequality in bipartite
single-qubit partition, which can be written as \cite{osb06prl}
\begin{equation}\label{1}
\mathcal{C}_{A_{1}|A_{2}A_{3}\cdots A_{n}}^{2}\ge
\mathcal{C}_{A_{1}A_{2}}^{2}+\mathcal{C}_{A_{1}A_{3}}^{2}+\ldots
+\mathcal{C}_{A_{1}A_{n}}^{2}.
\end{equation}
The entanglement between subsystems $A_{1}$ and $A_{2}A_{3}\ldots
A_{n}$ is quantified by $\mathcal{C}_{A_{1}|A_{2}A_{3}\ldots
A_{n}}^2(\rho_{A_{1}A_{2}A_{3}\cdots A_{n}})=\mbox{min} \sum_x p_x
\tau_{A_1}(\rho_{A_1}^{x})$, where the
$\tau_{A_1}(\rho_{A_1}^{x})=2[1-\mbox{tr}(\rho_{A_1}^{x})^2]$ is the
linear entropy \cite{san00pra,osb05pra}, and the minimum runs over
all the pure state decompositions. For the two-qubit quantum state
$\rho_{A_{i}A_{j}}$, its entanglement is analytically expressed as
$\mathcal{C}_{A_{i}A_{j}}^2=[\mbox{max}(0,
\sqrt{\lambda_1}-\sqrt{\lambda_2}-\sqrt{\lambda_3}-\sqrt{\lambda_4})]^2$,
with the decreasing nonnegative real numbers $\lambda_{i}$ being the
eigenvalues of the matrix
$\rho_{ij}(\sigma_y\otimes\sigma_y)\rho_{ij}^{\ast}(\sigma_y\otimes\sigma_y)$
\cite{woo98prl}. Based on the sum of the residual entanglements
$M_{A_i}=\mathcal{C}_{A_{i}|R(A_{i})}^2-\sum_j
\mathcal{C}_{A_iA_j}^2$, a multipartite entanglement measure for
pure states is introduced \cite{byw07pra,baw08pra}.

Now we analyze the multi-qubit entanglement distribution under
bipartite two-qubit partition. First, we consider a $2N$-qubit mixed
state $\rho_{A_{1}A_{1}^{\prime}A_{2}A_{2}^{\prime}\ldots
A_{n}A_{n}^{\prime}}$ with the reduced density matrix
$\rho_{A_{i}A_{i}^{\prime}}$ being a rank-2 quantum state. For this
quantum state, the following relations hold:
\begin{subequations}
\begin{eqnarray}
&&\mathcal{C}_{A_{1}A_{1}^{\prime}|A_{2}A_{2}^{\prime}\ldots
A_{n}A_{n}^{\prime}}^{2}\nonumber\\
&&\geq \sum_{i=2}^{n}
\mathcal{C}_{A_{1}A_{1}^{\prime}|A_{i}A_{i}^{\prime}}^{2}\\
&&\geq \sum_{i=2}^{n} \mathcal{C}_{A_{1}|A_{i}A_{i}^{\prime}}^{2}+
\sum_{i=2}^{n}\mathcal{C}_{A_{1}^{\prime}|A_{i}A_{i}^{\prime}}^{2}\\
&&\geq \sum_{i=2}^{n}( \mathcal{C}_{A_{1}A_{i}}^{2}+
\mathcal{C}_{A_{1}A_{i}^{\prime}}^{2}+\mathcal{C}_{A_{1}^{\prime}A_{i}}^{2}+
\mathcal{C}_{A_{1}^{\prime}A_{i}^{\prime}}^{2}).
\end{eqnarray}
\end{subequations}
In the derivation of the above inequalities, we have used the
property that $A_{i}A_{i}^{\prime}$ is equivalent to a single qubit
and the monogamy relation in Eq. (1). We here refer to the
inequalities (2a) and (2b) as the \emph{strong monogamy relations},
and the inequality (2c) as the \emph{weak monogamy relation}. In the
rank-2 case, we define the two-qubit residual entanglement as
\begin{equation}
M_{A_iA_{i}^{\prime}}(\rho_{A^{\otimes N}A^{\prime\otimes
N}})=\mathcal{C}_{A_iA_{i}^{\prime}|R(A_iA_{i}^{\prime})}^{2}-\sum
\mathcal{C}_{ij}^2,
\end{equation}
where $R(A_iA_{i}^{\prime})$ denotes the subset of qubits other than
$A_iA_{i}^{\prime}$, and $i$, $j$ in the sum represent the qubit in
the subsets $\{A_i, A_{i}^{\prime}\}$ and
$\{R(A_iA_{i}^{\prime})\}$, respectively. It is obvious that the
residual entanglement is zero when the $2N$-qubit state is separable
under the two-qubit partition. As a nontrivial example, we consider
the $2N$-qubit $W$ state, which can be written as
$\ket{W}_{2N}=\alpha_1 \ket{10\ldots 00}+\alpha_2 \ket{01\ldots
00}+\ldots+\alpha_{2n} \ket{00\ldots 01}$. For this quantum state,
we have
$\mathcal{C}_{A_{1}A_{1}^{\prime}|R(A_{1}A_{1}^{\prime})}^{2}=
4\sum_{i=1}^{2}\sum_{j=3}^{2n}|\alpha_i|^2 \cdot |\alpha_j|^2$ and
$\mathcal{C}_{ij}^{2}=4|\alpha_i|^2\cdot |\alpha_j|^2$. Then,
according to Eq. (3), the two-qubit residual entanglement is zero.
Since the square of the concurrence is a good entanglement measure
for two-qubit quantum state, the nonzero residual entanglement
$M_{A_{i}A_{i}^{\prime}}$ implies the existence of multipartite
entanglement.

While for the two-qubit partition of rank-3 and rank-4 cases, the
monogamy relation in Eq.(2) may not hold~\cite{note1}.

\section{Entanglement evolution in multipartite cavity-reservoir systems}
In Ref. \cite{clo08prl}, L\'{o}pez \emph{et al.} analyzed the
entanglement dynamics of two cavities interacting with independent
reservoirs. The initial quantum state of the composite system is
$\ket{\Phi_0}=(\alpha\ket{00}+\beta\ket{11})_{c_{1}c_{2}}\ket{00}
_{r_{1}r_{2}}$, where the two entangled cavity photons are in a
Bell-like state and their corresponding dissipation reservoirs are
in the vacuum states. The interaction Hamiltonian of a single cavity
and an $N$-mode reservoir is $H=\hbar \omega
a^{\dagger}a+\hbar\sum_{k=1}^{N}\omega_{k}
b_k^{\dagger}b_k+\hbar\sum_{k=1}^{N}g_{k}(ab_{k}^{\dagger}+b_{k}a^{\dagger})$.
Under the unitary evolution $U(H,t)=U_{c_{1}r_{1}}(H,t)\otimes
U_{c_{2}r_{2}}(H,t)$, the output state is given by
\begin{equation}
\ket{\Phi_t}=\alpha\ket{0000}_{c_{1}r_{1}c_{2}r_{2}}+\beta\ket{\phi_t}
_{c_{1}r_{1}}\ket{\phi_t}_{c_{2}r_{2}},
\end{equation}
where $\ket{\phi_t}=\xi(t)\ket{10}_{cr}+\chi(t)\ket{01}_{cr}$, and
the amplitudes $\xi(t)=\mbox{exp}(-\kappa t/2)$ and
$\chi(t)=[1-\mbox{exp}(-\kappa t)]^{1/2}$ in the large $N$ limit.
For this dynamical process, L\'{o}pez \emph{et al.} disclosed an
intrinsic connection between the ESD of the cavities and the ESB of
the reservoirs. However, it is not clear whether one can establish a
quantitative relation of the entanglements in different subsystems
in the process. Furthermore, it is still a subtle issue where the
entanglement really goes in the time window between the ESD and the
ESB.

We first show that an entanglement monogamy relation exists and
restricts the dynamical process of the multipartite systems. The
reduced density matrix of a single cavity with its reservoir is
$\rho_{c_{1}r_{1}}(t)=U_{c_{1}r_{1}}[\rho_{c_{1}r_{1}}(0)]U_{c_{1}r_{1}}^{\dagger}$,
where $\rho_{c_{1}r_{1}}(0)=|\alpha|^2\proj{00}+|\beta|^2\proj{10}$
is a rank-2 two-qubit state. Since the unitary operation does not
change the rank of the matrix, the $\rho_{c_{1}r_{1}}(t)$ is also a
rank-2 density matrix. Therefore, the entanglement monogamy
relations under the partition $c_{1}r_{1}|c_{2}r_{2}$ always hold in
the dynamical procedure. Particularly, we have
\begin{equation}
\mathcal{C}_{c_{1}r_{1}|c_{2}r_{2}}^2(t)\geq
\mathcal{C}_{c_{1}c_{2}}^2(t)+\mathcal{C}_{r_{1}r_{2}}^2(t)
+\mathcal{C}_{c_{1}r_{2}}^2(t)+\mathcal{C}_{c_{2}r_{1}}^2(t),
\end{equation}
where the two-qubit entanglements are
\begin{eqnarray}
\mathcal{C}_{c_{1}c_{2}}^2(t)&=&4[\mbox{max}(|\alpha\beta\xi^2|
-|\beta\xi\chi|^2,0)]^2,\nonumber\\
\mathcal{C}_{r_{1}r_{2}}^2(t)&=&4[\mbox{max}(|\alpha\beta\chi^2|
-|\beta\xi\chi|^2,0)]^2,\nonumber\\
\mathcal{C}_{c_{1}r_{2}}^2(t)&=&\mathcal{C}_{c_{2}r_{1}}^2(t)=4[\mbox{max}
(|\alpha\beta\xi\chi|-|\beta\xi\chi|^2,0)]^2.
\end{eqnarray}
Here, the bipartite entanglements are quantified by the square of
the concurrence rather than the concurrence in the analysis of
L\'{o}pez \emph{et al}. It should be emphasized that, once the
initial state is given, the bipartite entanglement
$\mathcal{C}_{c_{1}r_{1}|c_{2}r_{2}}^2(\Phi_t)=4|\alpha\beta|^2$ is
invariant in the entanglement evolution, where the invariance
property of entanglement under local unitary operations is used.

In Ref. \cite{clo08prl}, the multipartite entanglement is quantified
by the multipartite concurrence $C_{N}$ \cite{car04prl}. However,
$C_{N}$ is unable to characterize completely the genuine
multipartite entanglement.  For example, when the quantum state is a
tensor product of two Bell states, $C_{N}$ is nonzero. In this
paper, we consider the two-qubit residual entanglement
\begin{equation}
M_{c_{1}r_{1}}(\Phi_t)=\mathcal{C}_{c_1r_1|c_2r_2}^2(t)-\sum\mathcal{C}_{ij}^2(t),
\end{equation}
where $i\in \{c_1, r_1\}$ and $j\in\{c_2, r_2\}$. This quantity can
not only validate the monogamy relation, but also serve as an
\emph{indicator} of genuine multipartite entanglement in the
dynamical process. According to the expression of $\ket{\Phi_t}$ in
Eq. (4), one can deduce that all its three-tangles \cite{ckw00pra}
$\tau_3(\rho_{ijk})=0$, because $\rho_{ijk}$ can be written as the
mix of a $W$-state and a product state. Therefore, the nonzero
$M_{c_{1}r_{1}}(\Phi_t)$ indicates only the genuine four-qubit
entanglement. In Fig.1, we plot the residual entanglement
$M_{c_1r_1}$ as a function of the initial state amplitude $|\alpha|$
and the dissipation time $\kappa t$. For a given value of the
$\alpha$, the $M_{c_{1}r_{1}}(\kappa t)$ changes from zero to a
maximum value, and then decreases asymptotically to zero when
$\kappa t\rightarrow \infty$. Moreover, the maximum values of
$M_{c_{1}r_{1}}(\kappa t)$ occur in the time $\kappa t=\mbox{ln2}$
being independent of the amplitude $\alpha$. For all possible
$\alpha$, the maximum of the residual entanglement is
$M_{c_{1}r_{1}}(\alpha,\mbox{ln}2)=(13\sqrt{13}-19)/34\approx
0.81977$, where $|\alpha|=[(9+\sqrt{13})/34]^{1/2}\approx 0.60889$.

%%%%%%%%%%%%%%%%%%%%%%%%%%%%%%%%%%%%%%%%%%%%%%%%%%%%%%%%%%%%%%%%%%%%%%%%
\begin{figure}[t]
\begin{center}
\epsfig{figure=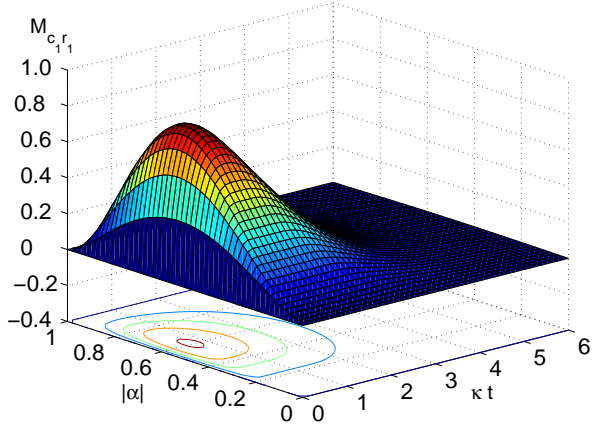,width=0.45\textwidth}
\end{center}
\caption{(Color online) Two-qubit residual entanglement
$M_{c_{1}r_{1}}(\Phi_t)$ vs the real parameters $|\alpha|$ and
$\kappa t$ in the entanglement evolution.}
\end{figure}
%%%%%%%%%%%%%%%%%%%%%%%%%%%%%%%%%%%%%%%%%%%%%%%%%%%%%%%%%%%%%%%%%%%%%%%%

%%%%%%%%%%%%%%%%%%%%%%%%%%%%%%%%%%%%%%%%%%%%%%%%%%%%%%%%%%%%%%%%%%%%%%%%
\begin{figure}[b]
\begin{center}
\epsfig{figure=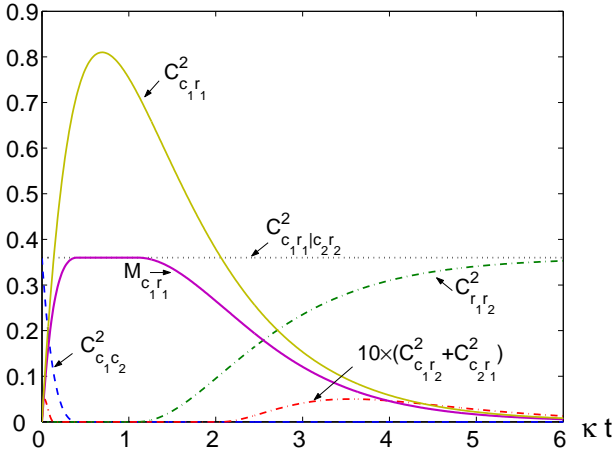,width=0.45\textwidth}
\end{center}
\caption{(Color online) The two-qubit residual entanglement
$M_{c_{1}r_{1}}$ (purple solid line) vs the time evolution parameter
$\kappa t$, in comparison to bipartite entanglements
$\mathcal{C}_{c_{1}c_{2}}^2$ (blue dashed line),
$\mathcal{C}_{r_{1}r_{2}}^2$ (green dot-dashed line),
$10(\mathcal{C}_{c_{1}r_{2}}^2+\mathcal{C}_{c_{2}r_{1}}^2)$ (red
dot-dashed line), $\mathcal{C}_{c_{1}r_{1}|c_{2}r_{2}}^2$ (black
doted line), and $\mathcal{C}_{c_{1}r_{1}}^2$ (yellow solid line) in
quantum state $\ket{\Psi_t}$ for which $\alpha=1/\sqrt{10}$
\cite{clo08prl}.}
\end{figure}
%%%%%%%%%%%%%%%%%%%%%%%%%%%%%%%%%%%%%%%%%%%%%%%%%%%%%%%%%%%%%%%%%%%%%%%%

Now, we look into the subtle issue where the initial entanglement
goes in the time interval when both cavity-cavity and
reservoir-reservoir entanglements are zero. We choose the initial
state parameter $\alpha=1/\sqrt{10}$ which is the same as that in
Ref. \cite{clo08prl}, and for this value, there is such a time
window. In Fig.2, we plot the two-qubit residual entanglement
$M_{c_{1}r_{1}}$ and related bipartite concurrences $\mathcal{C}^2$
against the parameter $\kappa t$.  The bipartite entanglement
$\mathcal{C}_{c_{1}r_{1}|c_{2}r_{2}}^2(\kappa t)$ in the process is
a conserved quantity ($=0.36$) and the monogamy relation in Eq. (5)
restricts the entanglement evolution. The two-qubit residual
entanglement $M_{c_{1}r_{1}}$ changes from zero to the maximum
$0.36$ in the time $[0,-\mbox{ln}(2/3)]$, then the value keeps
unchanged until $\kappa t=\mbox{ln}3$, and finally the
$M_{c_{1}r_{1}}$ decreases asymptotically to zero as the time
$\kappa t\rightarrow \infty$. This indicates that the genuine
multipartite entanglement is always involved in the dynamical
process. Particularly, in the plateau of $\kappa
t\in[-\mbox{ln}(2/3), \mbox{ln}3]$ where all the
$\mathcal{C}_{ij}^2(t)$ in Eq. (7) are zero, the initial
entanglement $\mathcal{C}_{c_{1}c_{2}}^2(0)=0.36$ transfers
completely to the genuine four-qubit entanglement in the composite
system (note that all the three-tangles are zero). In this region,
the $M_{c_{1}r_{1}}$ is just the
$\mathcal{C}_{c_{1}r_{1}|c_{2}r_{2}}^2$, and is entanglement
monotone, being able to characterize the genuine four-qubit
entanglement. For other initial state amplitudes satisfying
$|\alpha|<|\beta|/2$, there is also a plateau of
$M_{c_{1}r_{1}}(\kappa t)$ (see Fig.1) whose width and value are
$\kappa t_{\mbox{w}}=\mbox{ln} (|\beta/\alpha|-1)$ and
$M_{c_{1}r_{1}}=4|\alpha\beta|^2$, respectively. After a direct
comparison, we can get that the value is equal to the initial
cavity-cavity entanglement
($\mathcal{C}_{c_{1}c_{2}}^2(0)=4|\alpha\beta|^2$) and the width is
just the time window \cite{clo08prl} between the ESD of cavities and
the ESB of reservoirs. Here, according to Eq. (6), one can prove
further
$\mathcal{C}_{c_{1}r_{2}}^2(t)=\mathcal{C}_{c_{2}r_{1}}^2(t)=0$ in
the interval. Therefore, we conclude that \emph{the initial
entanglement evolves completely to the genuine four-partite
entanglement in the time window between the ESD of cavity subsystem
and the ESB of reservoir subsystem}.

We also wish to indicate that the nonzero
$\mathcal{C}_{c_1r_1}^2(t)$ in Fig.2 does not come from the initial
entanglement $\mathcal{C}_{c_1c_2}^2(0)$, but is generated by a
``local" unitary operation $U_{c_1r_1}(H,t)$ with the partition
$c_1r_1|c_2r_2$.

\section{Block-block entanglement versus genuine multipartite
entanglement} The multi-qubit entanglement property in the plateau
region is worthy of a further analysis. For the initial state with
$\alpha=1/\sqrt{10}$, the output state of the evolution can be
written as
\begin{equation}
\ket{\Psi_t}=\frac{1}{\sqrt{10}}\ket{0000}_{c_{1}r_{1}c_{2}r_{2}}
+\frac{3}{\sqrt{10}}\ket{\psi_t}_{c_{1}r_{1}}\ket{\psi_t}_{c_{2}r_{2}},
\end{equation}
where $\ket{\psi_t}=\xi(t)\ket{10}+\chi(t)\ket{01}$. Its genuine
four-qubit entanglement is evaluated in bipartite block-block form,
\emph{i.e.}, the entanglement measure
$M_{c_{1}r_{1}}(\Psi_t)=\mathcal{C}_{c_{1}r_{1}|c_{2}r_{2}}^2(\Psi_t)=0.36$
characterizes the \emph{genuine block-block entanglement} between
subsystems $c_{1}r_{1}$ and $c_{2}r_{2}$. The case for other
$\alpha$ with plateau region is similar.

Although the three-tangles $\tau_3(\rho_{ijk})$ and the related
$\mathcal{C}_{ij}^2$ in the plateau region are zero, the three-qubit
subsystems exhibit \emph{genuine qubit-block} entanglements and the
relation
\begin{equation}
\mathcal{C}_{c_{1}r_{1}|c_{2}r_{2}}^2(t)=\mathcal{C}_{c_{1}|c_{2}r_{2}}^2(t)
+\mathcal{C}_{r_{1}|c_{2}r_{2}}^2(t)
\end{equation}
holds, in which
$\mathcal{C}_{c_{1}|c_{2}r_{2}}^2(t)=4|\alpha\beta|^2|\xi(t)|^2$ and
$\mathcal{C}_{r_{1}|c_{2}r_{2}}^2(t)=4|\alpha\beta|^2|\chi(t)|^2$
being equivalent to the mixed state one-tangle \cite{osb06prl}. This
qubit-block entanglement is similar to that of mixed states in Refs.
\cite{loh06prl,byw08pra} which are entangled but without the
(two-qubit) concurrences and three-tangles. For this kind of
entanglement, our understanding is that it comes from the genuine
multipartite entanglement in its purified state \cite{byw08pra}.
Here, Eq. (9) actually presents for the first time a quantitative
relation for understanding the qubit-block entanglement, with a
schematic diagram being depicted in Fig.3.

%%%%%%%%%%%%%%%%%%%%%%%%%%%%%%%%%%%%%%%%%%%%%%%%%%%%%%%%%%%%%%%%%%%%%%%%
\begin{figure}
\begin{center}
\epsfig{figure=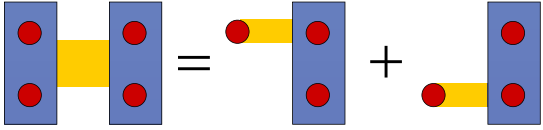,width=0.4\textwidth}
\end{center}
\caption{(Color online) The relation between the block-block
entanglement and qubit-block entanglement in the plateau region.}
\end{figure}
%%%%%%%%%%%%%%%%%%%%%%%%%%%%%%%%%%%%%%%%%%%%%%%%%%%%%%%%%%%%%%%%%%%%%%%%

\section{Discussion and conclusion}
Entanglement monogamy is a fundamental property of multipartite
entangled states. We argue that the violation of the monogamy
relations in Eq. (2) for higher rank cases is because the square of
the concurrence does not have the additivity, \emph{i.e.},
$\mathcal{C}_{A_{1}A_{1}^{\prime}|A_{2}A_{2}^{\prime}}^2\neq
\mathcal{C}_{A_{1}A_{2}}^2
+\mathcal{C}_{A_{1}^{\prime}A_{2}^{\prime}}^2$ for the tensor
product of two Bell states. The von Neumann entropy has this
additivity property, however, it has the negative residual
entanglement for multipartite systems \cite{wus00pra}. How to define
an additive entanglement measure with nonnegative residual
entanglement is still challenging.

The monogamy relations in Eq. (2) can be applied to other systems
\cite{guo09ept} only if the individual system-environment is in a
rank-2 quantum state and the evolution has a tensor structure
$U(H,t)=U_{S_1E_1}(H,t)\otimes U_{S_2E_2}(H,t)\otimes\ldots \otimes
U_{S_nE_n}(H,t)$. Moreover, based on this relation, one can derive
other useful monogamy inequality. For example, if the initial state
of a three cavity-reservoir composite system is
$\ket{\Psi_0}=(\alpha\ket{000}+\beta\ket{111})_c\ket{000}_r$ and the
individual cavity-reservoir interaction is the same as the previous
one, we can derive
\begin{equation}
\mathcal{C}_{c_{1}r_{1}|c_{2}r_{2}c_{3}r_{3}}^2(0)\geq\tau_3
(\rho_{c_{1}c_{2}c_{3}}(t))+\tau_{3}(\rho_{r_{1}r_{2}r_{3}}(t)),
\end{equation}
where
$\mathcal{C}_{c_{1}r_{1}|c_{2}r_{2}c_{3}r_{3}}^2(0)=4|\alpha\beta|^2$
gives an upper bound for the three-tangles in the entanglement
evolution.

In conclusion, we show that a monogamy relation restricts the
entanglement evolution of two cavities with individual reservoirs.
Moreover, based on the relation, we find the initial state
entanglement evolves completely to the genuine four-partite
entanglement in the time interval between the ESD of cavity-cavity
entanglement and the ESB of the reservoir-reservoir entanglement. In
addition, we give a quantitative relation between the block-block
entanglement and the qubit-block entanglement.

\section*{Acknowledgments}
The work was supported by the RGC of Hong Kong under Grants No.
HKU7051/06P, HKU7044/08P, and the URC fund of HKU. Y. K. B. was also
supported by the fund of Hebei Normal University and NSF-China Grant
10905016. M.Y.Y. was also supported by NSF-China Grant No. 60878059.


\begin{thebibliography}{99}
\bibitem{note1}     The counterexample for rank-3 case is the quantum state
                    $\ket{\Psi}_{A_{1}A_{1}^{\prime}A_{2}A_{3}}=(\sqrt{2}\ket{0010}+
                   \sqrt{2}\ket{0101}+\ket{1000}+\ket{1011})/\sqrt{6}$, for which the strong monogamy
                    relation in Eq. (2b) does not hold and has the form
                    $4/3<8/9+8/9$. The counterexample for rank-4 case is
                    the tensor product of two Bell states
                    $\ket{\Phi}_{A_{1}A_{1}^{\prime}A_{2}A_{2}^{\prime}}=
                    \ket{\psi^+}_{A_1A_2}\otimes\ket{\psi^+}_{A_{1}^{\prime}A_{2}^{\prime}}$,
                    for which neither the strong nor the weak monogamy relation is satisfied.
\end{thebibliography}
\end{document}